\documentclass[prc,twocolumn,tightenlines,showpacs]{revtex4}
\usepackage{graphicx}
\begin{document}

\title{Breakup of $^{17}$F on $^{208}$Pb near the Coulomb barrier}
\date{\today}

\author{J. F. Liang}
\author{J. R. Beene}
\author{A. Galindo-Uribarri}
\author{J. Gomez del Campo}
\author{C. J. Gross}
\author{P. A. Hausladen}
\author{P. E. Mueller}
\author{D. Shapira}
\author{D. W. Stracener}
\author{R. L. Varner}
\affiliation{Physics Division, Oak Ridge National Laboratory, Oak Ridge,
Tennessee 37831}
\author{J. D. Bierman}
\affiliation{Physics Department AD-51, Gonzaga University, Spokane, Washington
99258-0051}
\author{H. Esbensen}
\affiliation{Physics Division, Argonne National Laboratory, Argonne, Illinois
60439}
\author{Y. Larochelle}
\affiliation{Department of Physics and Astronomy, University of Tennessee,
Knoxville, Tennessee 37966}

\begin{abstract}
Angular distributions of oxygen produced in the breakup of $^{17}$F incident
on a $^{208}$Pb target have been measured around the grazing angle at beam
energies of 98 and
120 MeV. The data are dominated by the proton stripping mechanism and are well
reproduced by dynamical calculations. The measured breakup cross section is
approximately a factor of 3 less than that of fusion at 98 MeV. The influence
of breakup on fusion is discussed.
\end{abstract}


\maketitle

\section{introduction}
The study of nuclear reactions near the Coulomb barrier involving loosely
bound nuclei has received considerable attention in recent years. This is 
primarily driven by the advent of radioactive ion beams\cite{si01}.
It is frequently observed in stable
beam experiments that the subbarrier fusion cross sections are
enhanced over one-dimensional barrier penetration model predictions. The
enhancement can be described by channel couplings where the interplay of
the intrinsic degrees of freedom and reaction channels modify the
single barrier to multiple barriers\cite{da98}. The barriers appearing at lower
energies are responsible for the fusion enhancement. Breakup is a major
reaction channel in the scattering of loosely bound nuclei; this removal of
the incident flux would lead to fusion suppression\cite{hu92,ta93}.
On the other hand, the coupling to the breakup channel can change the barrier
distribution which could result in fusion enhancement\cite{da94}. Experimental
efforts have been put forward to study the influence of breakup on subbarrier
fusion\cite{si01}.

The fusion excitation functions of the neutron skin nucleus $^{6}$He on
$^{209}$Bi\cite{ko98a,ko98b,de98}
and $^{238}$U\cite{tr00} were measured and large subbarrier fusion enhancements
were observed in both cases. The breakup cross sections of
$^{6}$He on $^{209}$Bi
measured below the Coulomb barrier are orders of magnitude greater than
fusion\cite{ag00,ag01}. Measurements with stable $^{9}$Be,
which has a neutron
binding energy of 1.665 MeV, on $^{208}$Pb\cite{da99} and
$^{209}$Bi\cite{yo96,si98} found that the fusion
was not enhanced below the Coulomb barrier and was suppressed by about 30\%
above the barrier. In the $^{9}$Be+$^{208}$Pb reaction, the incomplete fusion
reaction $\alpha$+$^{208}$Pb following the breakup of $^{9}$Be into 
n+$\alpha$+$\alpha$ was observed. The suppression of fusion at energies above
the barrier was attributed to the projectile breakup. The barrier distribution
extracted from the fusion excitation functions is consistent with a single
barrier for
the $^{9}$Be+$^{208}$Pb and $^{9}$Be+$^{209}$Bi\cite{si02} even though
very large breakup yields were observed below the barrier\cite{si02}.
The fusion of $^{9}$Be+$^{209}$Bi was compared to that of $^{11}$Be
(a neutron halo nucleus with neutron binding energy of 0.504 MeV)+$^{209}$Bi.
At energies below the barrier the cross
sections were similar to that of $^{9}$Be+$^{209}$Bi whereas at energies
above the barrier the cross sections are significantly larger than the
predictions from
a coupled-channels calculation which takes into account the large rms
radius of $^{11}$Be. However, the precision of the data was not very good and
further measurements are required\cite{si02}.

On the proton rich side, fusion of a proton drip line nucleus, $^{17}$F,
with $^{208}$Pb was measured\cite{re98}.
The fusion excitation function is almost 
identical to that of $^{16}$O+$^{208}$Pb and $^{19}$F+$^{208}$Pb after
correcting for the Coulomb
barrier arising from the charge and size differences. There was no
enhancement or perhaps a small suppression of fusion below the barrier. 
It is noted that the loosely bound proton can be polarized in the
large Coulomb field of
the target in such a way that the proton is shielded by the core
and the breakup probability is reduced\cite{es96,es02}.
This paper reports the breakup of $^{17}$F on $^{208}$Pb
measured near the Coulomb barrier.

\section{experiment}
The experiment was carried out at the Holifield Radioactive Ion Beam Facility
(HRIBF) where the Isotope Separator On-Line (ISOL) technique was employed for
radioactive ion beam production. A 44 MeV deuteron beam from the Oak Ridge
Isochronous Cyclotron (ORIC) was incident on a fibrous hafnium oxide target
to produce
short-lived $^{17}$F by the $^{16}$O(d,n)$^{17}$F reaction\cite{we99}.
The reaction products were extracted from a closely coupled kinetic ejection
negative ion source\cite{al00}, mass analyzed, and accelerated
by the 25 MV tandem electrostatic accelerator. The $^{17}$O isobar
was removed from the accelerated beams by inserting an 80 $\mu$g/cm$^{2}$
carbon foil at the exit of the tandem accelerator
and selecting the 9$^{+}$ ions with a 90
degree analyzing magnet. The target used was a self-supporting $^{208}$Pb foil 
with a nominal thickness of 1.8 mg/cm$^{2}$. The reaction energies, 98 and 120
MeV, were calculated for the beams at the middle of the target by taking into
account the energy loss in the target. The beam intensity was measured
by detecting the secondary electrons generated during transmission of the beam
through a 10 $\mu$g/cm$^{2}$ carbon foil with a microchannel plate detector.
The average intensity was $1.5 \times 10^{6}$ and $8 \times 10^{6}$
$^{17}$F$^{9+}$/s for the 98 and 120 MeV reactions, respectively. The highest
intensity achieved was 10$^{7}$ ions/s for the 120 MeV beam.

The reaction products were detected by a $\Delta E$-$E$ telescope composed of
a 29 $\mu$m Si detector mounted in
front of a 1000 $\mu$m double-sided silicon strip detector (DSSD). The area
of the Si detector and the DSSD is 5$\times$5 cm$^{2}$. The DSSD, which has 16
vertical and 16 horizontal strips, was placed near the grazing angle at
10.5 cm from the target and
symmetric with respect to the horizontal plane,
{\em i.e.} half of the detector above and half of the detector below the plane.
At backward angles, the variation of 
scattering angle for pixels on a vertical strip is small. Events in pixels on 
the same vertical strip can be summed to increase statistics. At
forward angles, pixels on the same vertical strip have to be divided into
two groups, 8 middle pixels and 8 outer pixels,
in order to keep the angular spread similar to the backward angles 
($\simeq 2^{\circ}$). The uniformity of the Si detector was determined
by measuring elastic
scattering at forward angles. The position of the elastically scattered
particles penetrating the detector was obtained from the overlap of the
horizontal and vertical strips of the DSSD. The energy loss of the elastically 
scattered
particles in the 256 (16$\times$16) pixels was compared to kinematics
and stopping power calculations. 
Two 50 mm$^{2}$ Si surface barrier detectors placed at 10$^{\circ}$ on either
side of the beam were used to monitor the beam position and for normalization
between runs.

\begin{figure}[h]
\includegraphics[width=2.25in, angle=90]{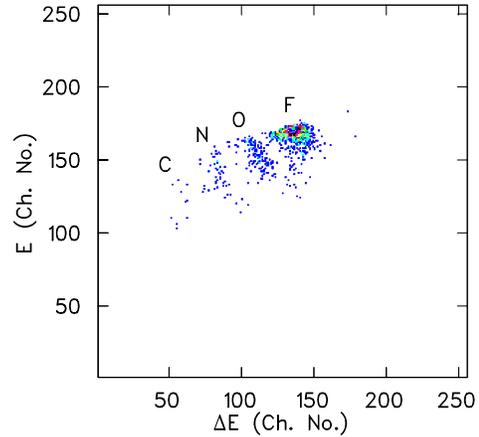}
\caption{\label{fg:ede}Histogram of E versus $\Delta$ E for reaction
products produced in 120 MeV $^{17}$F+$^{208}$Pb measured at 
$\theta_{lab}=65^{\circ}$ by summing events in a vertical strip of the DSSD.}
\end{figure}

An $E$ versus $\Delta E$ plot for the 120 MeV $^{17}$F-induced reaction is
displayed in Fig. \ref{fg:ede}.
It is obtained by summing events in the pixels of one vertical strip at
$\theta_{lab} = 65^{\circ}$. The energy loss, $\Delta E$,
was corrected for the nonuniformity of the Si detector. A group of oxygen
events can be clearly identified and is well separated from the
elastically scattered $^{17}$F.
\section{result and discussion}

\begin{figure}[h]
\includegraphics[width=2.0in, angle=90]{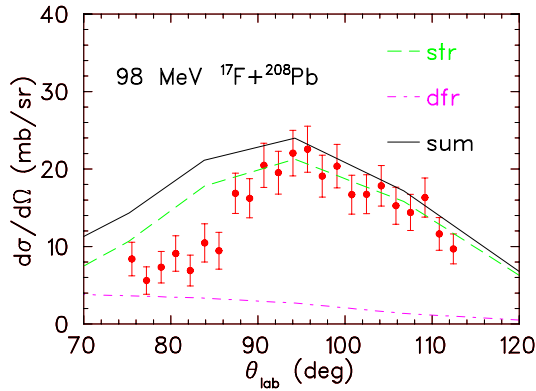}
\caption{\label{fg:o16e98}Angular distribution of oxygen produced from 98 MeV
$^{17}$F+$^{208}$Pb. The calculated stripping and diffraction breakup are
shown by the dashed and dash-dotted curves, respectively. The solid curve
is for the sum of the two.}
\end{figure}

\begin{figure}[h]
\includegraphics[width=2.0in, angle=90]{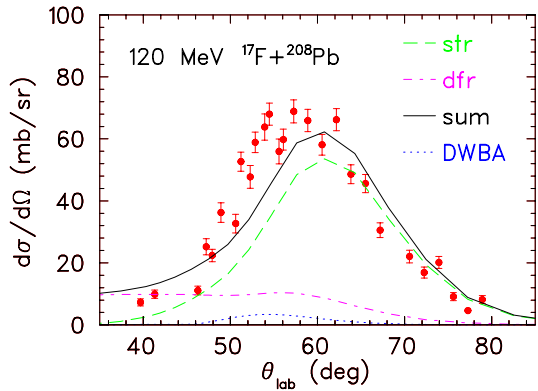}
\caption{\label{fg:o16e120}Angular distribution of oxygen produced from 120
MeV $^{17}$F+$^{208}$Pb. The calculated stripping and diffraction breakup are
shown by the dashed and dash-dotted curves, respectively. The solid curve
is for the sum of the two. The results of one-step DWBA transfer calculations
are shown by the dotted curve.}
\end{figure}

The angular distributions of oxygen produced in $^{17}$F+$^{208}$Pb collisions
at 98 and 120 MeV are shown in Fig. \ref{fg:o16e98} and 
Fig. \ref{fg:o16e120}, respectively. The angular distributions are bell shaped
and have a peak near the grazing angle. Since the $\Delta E$-$E$ telescope is 
not able to resolve mass, calculations were performed to estimate contributions
of reactions leading to oxygen isotopes other than $^{16}$O.
The charge exchange reaction ($^{17}$F,$^{17}$O) has a Q-value of --0.11 MeV.
Two-step Distorted Wave Born Approximation (DWBA) calculations using the code
{\sc fresco}\cite{th88} were performed
to estimate the contribution of this reaction. Sequential single-nucleon
transfer reactions, $^{17}$F $\rightarrow$ $^{16}$O $\rightarrow$ $^{17}$O
and $^{17}$F $\rightarrow$ $^{18}$F $\rightarrow$ $^{17}$O, were calculated.
In the calculations, transfer to excited states in the projectile- and 
target-like nuclei were included. States which have large spectroscopic factors
measured in light ion transfer reactions or large cross sections calculated in 
one-step single-nucleon transfer reactions were
selected for the two-step DWBA calculations.
Table~\ref{tb:qxfr} presents the states included in the calculations.
In these calculations, the spectroscopic factors were set to
1.0 to estimate the magnitude of the yields.
The shape of the calculated ($^{17}$F,$^{17}$O) angular distribution at
E$_{lab}$ = 120 MeV is similar to the measured angular distribution
and has a peak at $\theta_{lab}=58^{\circ}$.
However, the calculated peak cross section is 0.0028 mb/sr which is several
orders of
magnitude less than the measured value. 
Although reactions leading to $^{18}$O and $^{207}$Bi in the exit channel have
positive Q-values, they cannot occur by simple single-step transfer processes.
Therefore, the cross sections are expected to be smaller than that for 
one-proton transfer\cite{ko74}.
The results of DWBA calculations for one-step proton transfer
$^{208}$Pb($^{17}$F,$^{16}$O)$^{209}$Bi at 120 MeV are shown by the dotted
curve in Fig.~\ref{fg:o16e120}. One proton transfer to the six lowest
single particle states in $^{209}$Bi was calculated by the code
{\sc Ptolemy}\cite{ma78} with the spectroscopic factors set to 1.0. 
It can be seen that neither
one-proton transfer nor charge exchange can account for the measured oxygen
angular distribution. Since the direct charge exchange is orders of magnitude 
smaller than that of one-nucleon transfer\cite{du74},
its contribution to the data can be safely ignored.
\begin{table}
\caption{\label{tb:qxfr}States included in calculations of the
($^{17}$F,$^{17}$O) reaction by successive nucleon transfer.} 
\begin{ruledtabular}
\begin{tabular}{ccc}
Nucleus    & E$^{*}$ (MeV) & $J^{\pi}$          \\ \hline
$^{16}$O   & 0.0           & $0^{+}$            \\ \hline
$^{18}$F   & 0.0           & $1^{+}$            \\
           & 0.937         & $3^{+}$            \\ \hline
$^{17}$O   & 0.0           & $\frac{5}{2}^{+}$  \\
           & 0.871         & $\frac{1}{2}^{+}$  \\ \hline
$^{209}$Bi & 0.0           & $\frac{9}{2}^{-}$  \\
           & 0.896         & $\frac{7}{2}^{-}$  \\
           & 1.609         & $\frac{13}{2}^{+}$ \\ \hline
$^{207}$Pb & 0.0           & $\frac{1}{2}^{-}$  \\ \hline
$^{208}$Bi & 0.0           & $5^{+}$            \\
           & 0.063         & $4^{+}$            \\
           & 0.937         & $3^{+}$            \\
           & 1.034         & $4^{+}$            \\
\end{tabular}
\end{ruledtabular}
\end{table}

\begin{figure}[h]
\includegraphics[width=2.0in, angle=90]{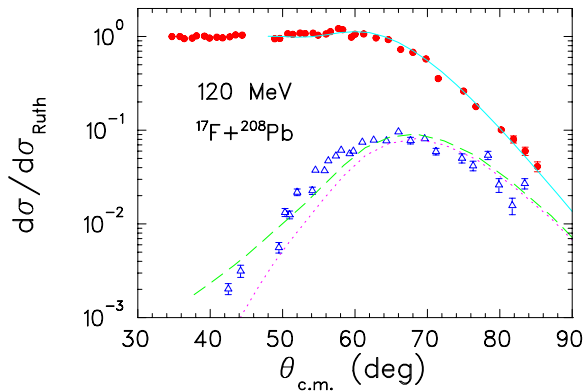}
\caption{\label{fg:elastic}Angular distribution of elastic scattering (filled
circles) in 120 MeV $^{17}$F+$^{208}$Pb. The result of an optical model
fit to the data is shown by the solid curve. The angular distribution of
oxygen produced in the same reaction is presented for comparison (open
triangles). The calculated stripping is shown by the dotted curve and the sum
of stripping and diffraction breakup is shown by the dashed curve.}
\end{figure}

The measured angular distributions are compared to results of dynamical
calculations where the relative motion of the proton and the $^{16}$O core is
described quantum mechanically by solving the time dependent Schr\"{o}dinger 
equation for the two-body breakup in the Coulomb and nuclear
fields from the target nucleus\cite{es95}.
It has been shown that calculations of this
kind are suitable for energies near the Coulomb barrier\cite{es99}.
The time evolution
of the projectile wave function was calculated to obtain the angular
distribution of $^{16}$O from the $^{17}$F$\rightarrow$$^{16}$O+p reaction.
The breakup angular distribution is obtained by multiplying the breakup
probability calculated as a function of impact parameter by a fit to the
measured elastic scattering
cross section at the corresponding Rutherford scattering angle.
The measured angular distribution of elastic
scattering at 120 MeV is shown in Fig.~\ref{fg:elastic}. Since the angular
resolution is $\sim 2^{\circ}$ and the cross sections fall off exponentially at
large angles, exponential functions fitted between two adjacent data points
were used to calculate the weighted average of each data point
at large angles.
The solid
curve shows an optical model fit to the elastic scattering, which is used
in converting calculated breakup probabilities into an angular distribution.
The resulting breakup cross section is shown by the dashed curve in 
Fig.~\ref{fg:elastic}. The dotted curve is the separate contribution 
from stripping, which dominates the breakup and is in fairly good
agreement with the measurement shown by the open triangles.

\begin{figure}[h]
\includegraphics[width=2.0in, angle=90]{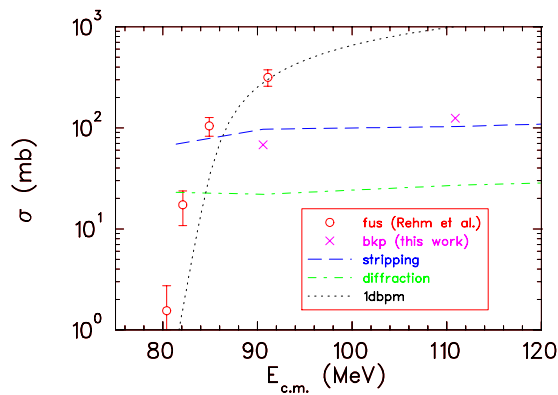}
\caption{\label{fg:fus}Comparisons of the fusion excitation function
($\circ$)
of $^{17}$F+$^{208}$Pb measured by Rehm {\em et al.}\protect\cite{re98}
and angle-integrated breakup cross sections measured in this work ($\times$).
The dotted curve is the one-dimensional barrier penetration model
prediction and the dashed and dash-dotted curves are for stripping and
diffraction breakup, respectively, predicted by dynamical calculations.}
\end{figure}

The calculated breakup cross section is compared to
the measured angular distribution of oxygen fragments
in Fig. \ref{fg:o16e98} and Fig.~\ref{fg:o16e120}, for 98 and 120 MeV,
respectively.
The  dashed curve is the stripping and the dash-dotted curves is
the diffraction dissociation, and their sum is shown by the solid curve.
It can be seen that the measured angular
distribution is predominantly due to the stripping breakup reaction.
The agreement between the data and calculations is very good for the 98~MeV
measurement. For the 120~MeV reaction, the measured
angular distribution is shifted slightly forward compared to the calculated
distribution but the total cross sections are in good agreement (see
Fig.~\ref{fg:fus}).

The angular distributions  of oxygen were fit to a Gaussian function to
obtain angle integrated breakup cross sections. Fig. \ref{fg:fus} displays
the fusion excitation function measured by Rehm {\em et al.}\cite{re98} and the
breakup cross sections measured in this work. The calculated
diffraction and stripping breakup are shown by the dash-dotted and dashed
curves,
respectively. As it was seen in the angular distribution of oxygen fragments
that the measured breakup is dominated by proton stripping,
the angle integrated breakup cross sections are in good agreement with the
calculated stripping cross sections. Near the barrier, the diffraction
breakup is a factor of 3 less than stripping which, in turn, is about
a factor of 3 less than fusion.

\begin{figure}[h]
\includegraphics[width=2.0in, angle=90]{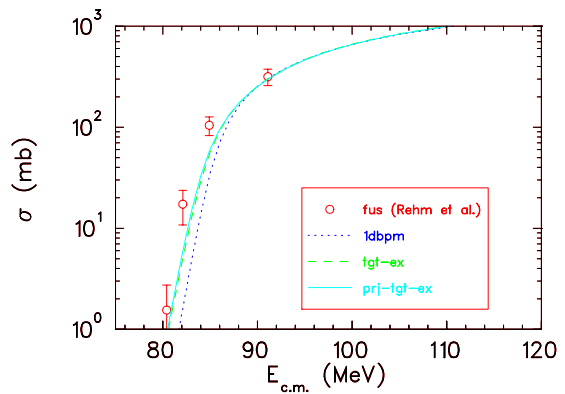}
\caption{\label{fg:ccex}Fusion excitation function for $^{17}$F+$^{208}$Pb
predicted by a one-dimensional barrier penetration model (dotted curve),
coupled-channels calculations taking into account target excitations (dashed
curve), and projectile and target excitations (solid curve) as described
in the text.}
\end{figure}

The result of a one-dimensional barrier penetration model calculation for 
$^{17}$F+$^{208}$Pb using the code {\sc ccmod}\cite{da92} is shown in
Fig. \ref{fg:fus} and Fig.~\ref{fg:ccex} by the dotted curve. The barrier
potential
parameters, $V_{0}=235.5$~MeV, $r_{0}=1.1$~fm, and $a=0.65$~fm, were
taken from the analysis of fusion measurements of a neighboring system,
$^{16}$O+$^{208}$Pb\cite{da97}, since the excitation function is almost
identical to
that of $^{17}$F+$^{208}$Pb after correcting for the Coulomb barrier.
As can be seen, the calculation underpredicts the cross sections at
subbarrier energies. Coupled-channels calculations were performed with the
code {\sc ccmod} using
procedures employed for analysis of the $^{16}$O+$^{208}$Pb measurement
in Ref.~\cite{da97}. It is well
established that inelastic excitations of the projectile and target can
contribute to subbarrier fusion enhancement. In many cases, coupling to
inelastic excitation channels can account for the enhanced fusion rates.
Calculations including the excitation of $^{208}$Pb to the lowest 2$^{+}$,
3$^{-}$, and 5$^{-}$ states were carried out. Furthermore,
it was found in the analysis of the $^{16}$O+$^{208}$Pb data that the
coupling of double-phonon excitations is essential for reproducing the barrier
distribution. The coupling of two-phonon states $3^{-}\otimes 3^{-}$ in
the harmonic limit and all the resulting cross coupling terms, {\em e.g.} 
$3^{-}\otimes 5^{-}$ were considered in the calculations. The result is shown
by the dashed curve in Fig.~\ref{fg:ccex}. The calculation still underpredicts
the measurement. The first excited state of $^{17}$F is bound by 105~keV and
can be excited from the ground state with a large B(E2) value,
B(E2)$\downarrow$~=~63.4~e$^{2}$fm$^{4}$\cite{adndt79,br82}. The results of
coupled-channels calculations 
including the excitation of $^{17}$F is shown by the solid curve in
Fig.~\ref{fg:ccex}. It can be seen that the increase in the subbarrier cross
sections is very small when this projectile excitation is included
and the coupled-channels calculations still
underpredict the subbarrier cross sections.

\begin{figure}[h]
\includegraphics[width=2.00in, angle=90]{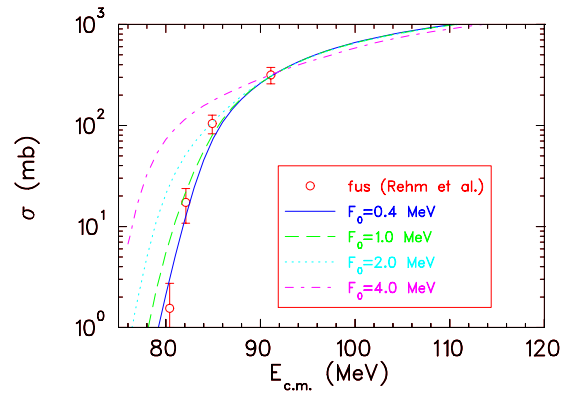}
\caption{\label{fg:ccxfr}Results of coupled-channels calculations including
inelastic excitations and transfer in $^{17}$F+$^{208}$Pb. The results for
transfer coupling constant $F_{0}$= 0.4, 1.0, 2.0, and 4.0 MeV are shown by the
solid, dashed, dotted, and dash-dotted curves, respectively.}
\end{figure}

In the fusion of $^{16}$O+$^{208}$Pb, the excitation function was reproduced
by calculations coupling to the inelastic excitation channels only. Since all
the
Q-values for neutron transfer are negative, it is not necessary to consider
transfer in the coupled-channels calculations. The neutron transfer Q-values
in $^{17}$F+$^{208}$Pb are positive for up to six-neutron pickup. In
particular, the two- and four-neutron transfer have Q-values greater than
5~MeV. This can be compared to $^{40}$Ca+$^{90}$Zr and $^{40}$Ca+$^{96}$Zr
where very large subbarrier fusion enhancement was observed in the
latter\cite{ti98}.
Coupling to the inelastic excitations of projectile and target only reproduces
the $^{40}$Ca+$^{90}$Zr measurement. There are still large discrepancies
between the measured cross sections of $^{40}$Ca+$^{96}$Zr and the
coupled-channels calculations. The major difference in the two reactions is
neutron transfer. The Q-values for multi-neutron transfer are negative
in $^{40}$Ca+$^{90}$Zr but positive in $^{40}$Ca+$^{96}$Zr. Measurements of
transfer near the barrier found large cross sections for the $^{96}$Zr
target\cite{mo02}. The influence of transfer on fusion is demonstrated in
a semiclassical model calculation\cite{mo02,wi94,wi95} where
the fusion cross sections as well as transfer for $^{40}$Ca+$^{96}$Zr are
reproduced. Fig.~\ref{fg:ccxfr} presents the results of coupled-channels
calculations including the inelastic excitations discussed above and nucleon
transfer.
The transfer is treated approximately in the code {\sc ccmod}, therefore
only qualitative comparison can be made here. The transfer form factor is
given by
\[ F(r) = \frac{F_{0}}{\sqrt{4\pi}}
exp\left[-\frac{(r-R_{1}-R_{2})}{a}\right]
{\rm MeV}, \]
where $F_{0}$ is the coupling constant, $R_{1,2}$ are the nuclear radii,
and $a$=1.2 fm is the diffuseness parameter. The solid, dashed, dotted, and
dash-dotted curves are for coupling
constant $F_{0}$ = 0.4, 1.0, 2.0, and 4.0 MeV, respectively. To simplify the
calculation, 
three channels: one-proton stripping, one-neutron pickup, and two-neutron
pickup, were included. The fusion excitation function can be reasonably
reproduced with $F_{0}$ set between 0.4 and 1 MeV. However,
it is noted that the
quantity $F_{0}$ can be as large as 3 or 4 MeV depending on the transferred
angular momentum and the orbitals occupied by the transferred
nucleons\cite{po83,da92,ti98}. Based on the calculations presented in
Fig.~\ref{fg:ccxfr}, the fusion excitation function can be reproduced
by including transfer of up to two nucleons in the calculations 
with the coupling constant $F_{0} \leq 1$ MeV. If channels of transferring more
than two nucleons are included and $F_{0} > 1$ MeV is used, the calculation
will
overpredict the measured cross sections, {\em i.e.} the fusion is suppressed
below the barrier. To better account for the influence of transfer on fusion
in $^{17}$F+$^{208}$Pb, measurements of multi-nucleon transfer and more
sophisticated
model calculations such as in Ref.~\cite{mo02} are required. Up to now, one
of the reactions which has not been considered in the
calculations is breakup. The simplified coupled-channels
code used here cannot treat breakup rigorously. Nevertheless, it is
conceivable that if fusion is suppressed, breakup can be responsible.
  
\begin{figure}[h]
\includegraphics[width=2.0in, angle=90]{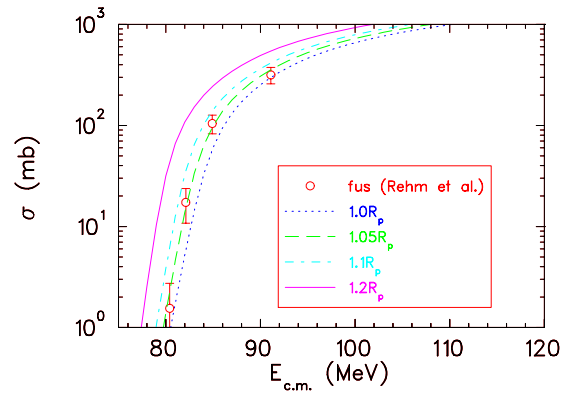}
\caption{\label{fg:radius}Results of coupled-channels calculations including
only inelastic excitations (dotted curve) and the radius of the projectile
increased by 5~\% (dashed curve), 10~\% (dash-dotted curve), and 20~\%
(solid curve).}
\end{figure}

It has been reported that
the $^{17}$F has a large rms radius, $<r>_{rms}=3.7$ fm\cite{mo97}.
In the coupled-channels calculations, the
nuclear radius is given by $r_{0}A^{1/3}$ where $r_{0}$=1.1 fm is the radius
parameter and $A$ is the mass
number. The effect of the large rms radius of $^{17}$F was
not accounted for in the previous calculations. To explore these effects,
the radius parameter of the projectile was adjusted in the
calculations. Since the treatment of coupling to transfer degrees of
freedom has large uncertainties introduced by the coupling constant 
$F_{0}$, only inelastic excitations were included. In
Fig.~\ref{fg:radius}, the dashed, dash-dotted, and solid
curves are for increasing the radius of $^{17}$F by 5~\%, 10~\%, and 20~\%,
respectively. It can be seen that the fusion excitation function can be
well reproduced by increasing the radius of $^{17}$F by 5~\% whereas
increasing the radius by 10~\% results in a calculated cross section that
exceeds
the measurements. If transfer channels were included, the discrepancy would be
larger. This further suggests that the fusion of $^{17}$F and $^{208}$Pb
may be suppressed.

In the $^{9}$Be+$^{208}$Pb reaction, complete fusion was found to be suppressed
at energies above the barrier. However, the sum of complete and
incomplete fusion agrees with a coupled-channels calculation.
The discrepancy between the measured complete fusion cross sections and
the coupled-channels prediction is attributed to the breakup of
$^{9}$Be\cite{da99}. The incomplete fusion arises from
$^{9}$Be breaking up into two $\alpha$ particles and a neutron, and
subsequently an $\alpha$ particle fuses with the target.
The fusion measurements in Ref.~\cite{re98} are made by detecting fission
fragments, and therefore probably determine total fusion-like cross section
(complete + incomplete fusion) rather than the complete fusion cross section
since the
incomplete fusion reaction, $^{16}$O+$^{208}$Pb, produces fission events
very similar to those of the complete
fusion reaction. Consequently, it is
not known whether the complete fusion of $^{17}$F+$^{208}$Pb is suppressed
above the barrier. Measurements of $^{17}$F on $^{208}$Pb at
10 MeV/nucleon
showed that it is necessary to consider core absorption ($^{16}$O absorbed
by $^{208}$Pb) in the dynamical calculation to reproduce the measured
diffraction breakup yield\cite{li02}. The dynamical calculations presented in
Fig.~\ref{fg:o16e98} and Fig.~\ref{fg:o16e120} also include core absorption.
It is expected that incomplete fusion is present in $^{17}$F+$^{208}$Pb.
A coincidence measurement of the breakup proton and the
fission fragments are required to identify the incomplete fusion reaction.

The predicted diffraction breakup seems too small to influence fusion
significantly. However, the stripping breakup yield is about one third of
fusion.
The energy dependence of this reaction predicted by the dynamical calculation
is presented by the dashed curve in Fig.~\ref{fg:fus}. The calculated stripping 
cross section exceeds that for fusion below the barrier. The measured 
breakup of $^{6,7}$Li and $^{9}$Be in the vicinity of the barrier in
$^{6,7}$Li+$^{208}$Pb\cite{ke00} and $^{9}$Be+$^{209}$Bi\cite{si02},
respectively, shows similar behavior. 
The analysis of elastic scattering of $^{6}$Li+$^{208}$Pb in Ref.~\cite{ke94}
shows that the imaginary
potential increases as the energy decreases below the barrier and the
threshold anomaly\cite{na85} disappears. Because of
this strong absorption, the enhancement of fusion at low energies should be
small and the breakup reaction is expected to be strong\cite{ke94}.
The measured fusion yields for $^{17}$F+$^{208}$Pb
were not enhanced and perhaps even slightly suppressed below the barrier.
It is conceivable that strong absorption exists resulting in large
stripping breakup which
removes $^{17}$F from the fusion channel. It would be interesting to
measure the elastic scattering and study the energy dependence of the
interaction potentials.

Large subbarrier fusion enhancement and transfer/breakup were observed in 
the $^{6}$He+$^{209}$Bi reaction\cite{ko98b,ag00}. Analysis of elastic
scattering indicated an absence of the threshold
anomaly\cite{ag01}. In this case, the strong absorption may not enhance fusion
much but may contribute mostly to transfer/breakup, as pointed out in
Ref.~\cite{ke94}. The neutron binding energy of
$^{6}$He is fairly low.  As suggested in
Ref.~\cite{ko98b}, the large subbarrier fusion enhancement may arise
from neutron flow since the threshold barrier correlates with
neutron binding energies\cite{st90}. In the $^{17}$F+$^{208}$Pb reaction,
the proton binding energy is very low but the proton flow must be strongly
suppressed
because of the Coulomb barrier. Therefore, the behavior of fusion below
the barrier is different from that of $^{6}$He+$^{209}$Bi. 

\section{summary and conclusion}

The breakup of 98 and 120 MeV $^{17}$F on $^{208}$Pb was measured by detecting
oxygen in singles. The angular distributions of oxygen are well
reproduced by dynamical calculations and found to be dominated by stripping
breakup. Near the barrier, the angle integrated stripping cross
section is about 30\% of that of fusion.
It has been shown in the analysis of $^{6}$Li+$^{208}$Pb elastic scattering
that the imaginary
potential continues to be large below the barrier. In this case, the breakup
yields are large but fusion is not much enhanced because the threshold
anomaly is absent. This may explain why a subbarrier fusion enhancement
was not observed in $^{17}$F+$^{208}$Pb.
Simplified coupled-channels calculations were performed to explore
the effects of coupling to both inelastic excitations and transfer degrees of
freedom on fusion. Furthermore, the radius of $^{17}$F was adjusted in the
calculations to study the change in the fusion excitation function.
The results suggest that fusion may be suppressed at energies below the
barrier. In contrast, large subbarrier fusion enhancements were observed for
fusion of the neutron skin nucleus $^{6}$He on $^{209}$Bi and $^{238}$U.
Further experiments are required to examine whether the differences observed
between the $^{6}$He- and $^{17}$F- induced fusion are due to breakup or other
reaction mechanisms. 
Measurements using neutron halo nuclei, such as $^{11}$Be and 
$^{11}$Li, and proton halo nuclei, such as $^{8}$B and $^{26}$P, would provide 
useful additional information.

\begin{acknowledgments}
We are grateful to F. M. Nunes and J. A. Tostevin for valuable discussions.
We would like to thank M. Dasgupta for discussion of the coupled-channels
code {\sc ccmod} and the analysis of their data.
Research at the Oak Ridge National Laboratory is 
supported by the U.S. Department of Energy under contract DE-AC05-00OR22725 
with UT-Battelle, LLC. One of us (H.E.) was
supported by the U.S. Department of Energy, Nuclear Physics Division, under
Contract No. W-31-109-ENG-38.
\end{acknowledgments}

\end{document}